\documentclass{aipproc}
\layoutstyle{6x9}
\usepackage{amssymb}

\newcommand{\pdd}{P.~Danielewicz}
\newcommand{\etal}{{\em et~al.}}              

\newcommand{\plb}[1]{Phys.~Lett.~B {#1}}
\newcommand{\prl}[1]{Phys.~Rev.~Lett.~{#1}}
\newcommand{\npa}[1]{Nucl.~Phys.~{A#1}}
\newcommand{\prc}[1]{Phys.~Rev.~C {#1}}

\newcommand{\beq}{\begin{equation}}
\newcommand{\eeq}{\end{equation}}
\newcommand{\bea}{\begin{eqnarray}}
\newcommand{\eea}{\end{eqnarray}}

\begin{document}

\title
      {Nuclear Equation of State}




\author{Pawe\l\  Danielewicz}{
  address={
National Superconducting Cyclotron Laboratory and
 Department of Physics and Astronomy,
Michigan State University, East Lansing, MI 48824, USA,\\ and
Gesellschaft f\"ur Schwerionenforschung mbH, D-64291 Darmstadt,
Germany},
  email={danielewicz@nscl.msu.edu}
}

\copyrightyear  {2001}

\begin{abstract}
Nuclear equation of state plays an important role in the
evolution of the Universe, in supernova explosions and, thus,
in the production of heavy elements,
and in stability of neutron
stars.  The equation constrains the two- and three-nucleon
interactions and the quantum chromodynamics in
nonperturbative regime. Despite the importance of the equation,
though, its features had remained fairly obscure.  The talk
reviews new results on the equation of state from measurements
of giant nuclear oscillations and from studies of
particle emission in central collisions of heavy nuclei.
\end{abstract}

\date{\today}

\maketitle

\section{Introduction}

An equation of state (EOS) is a nontrivial relation between
thermodynamic variables characterizing a medium.  While the
term is used in its singular form in nuclear physics, actually
different relations are of interest, such as between pressure
$p$ and baryon density $\rho$ and temperature $T$, $p(\rho,T)$,
or chemical
potential $\mu$ and $T$, $p(\mu,T)$, between energy density $e$
and
$\rho$ and $T$, $e(\rho,T)$, etc.  Some of the relations are
fundamental under certain conditions,
i.e.\ all other relations may be derived from them
(such as from
$e(\rho)$ at $T=0$).

The nuclear EOS is of interest because it affects the fate of
the
Universe at times $t \gtrsim 1 \, \mu$s from the Big Bang
and because its features are behind the supernova explosions.
Moreover, its features ensure the stability of neutron
stars.  Through
its effects on the evolution of the Universe, on supernovae
explosions, and on neutron-star collisions, the EOS affects
nucleosynthesis.  Moreover, the EOS
impacts central reactions of heavy nuclei.  Finally, the
form of the EOS constraints hadronic interactions and
the nonperturbative quantum chromodynamics (QCD).

\section{Importance of EOS}

\begin{figure}
{\includegraphics[angle=0,height=2.35in,
width=.50\linewidth]{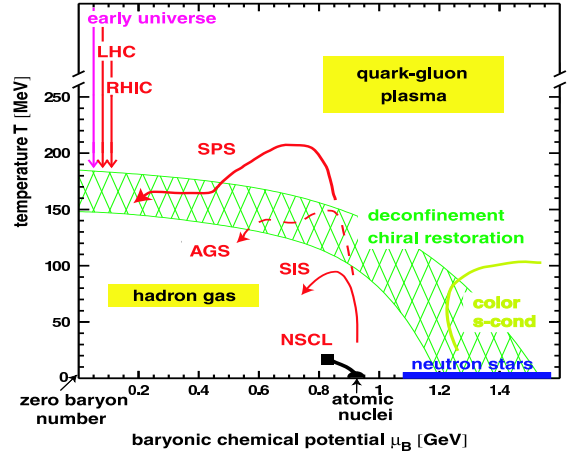}}
\caption{Strongly interacting matter in the $\mu - T$ plane,
after~\cite{sta99}$^*$.
}
\label{muT}
\end{figure}

\begin{figure}
{\includegraphics[angle=0,
width=.40\linewidth]{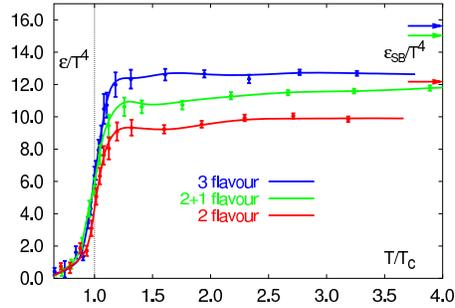}}
\caption{Energy in baryonless matter vs $T$, from calculations of
Ref.~\protect\cite{kar01}.}
\label{eT}
\end{figure}

Different regimes for the strongly interacting are conveniently
assessed in the $\mu-T$ plane, see Fig~\ref{muT}.  Along the
$T=0$ axis, at $\mu \approx 930$~MeV, we have the matter in
heavy nuclei.  The matter in the interior of neutron stars
corresponds to higher chemical potentials, in combinations with
low temperatures.
The matter in the early Universe evolved along the temperature
axis, at low baryon number content, and thus at low $\mu$.
Different regions of the plane are explored at
different accelerators.  In the early Universe and likely at
the higher-energy accelerators,
the
matter crosses the transition between the
hadronic matter and quark-gluon plasma.
The transition is observed in numerical
lattice QCD calculations as a rapid change in energy density in
the temperature region of $T_c \sim 170$~MeV, cf.\
Fig.~\ref{eT}.  The~numerical calculations are carried out on a
lattice of a finite size and it can be difficult to establish
whether one deals just with a transitional behavior or with a
phase
transition and, if so, of what order.  Whether or not there is
a first-order phase transition is of importance for the early
Universe.

\subsection{Early Universe}

\begin{figure}
{\includegraphics[angle=0,height=1.7in,
width=.46\linewidth]{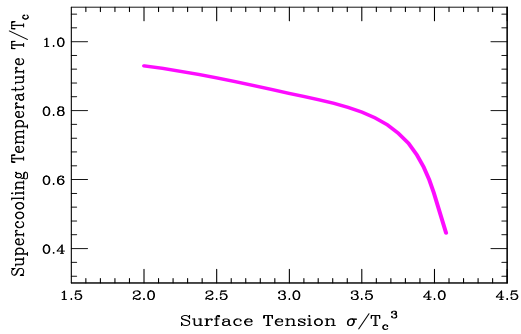} }
\caption{Supercooling tension for the confinement phase
transition, after \protect\cite{kam00}.}
\label{Ts}
\end{figure}

\begin{figure}
\parbox{1.40in}
{\vspace*{0.35in}
\includegraphics[angle=0,
width=1.\linewidth]{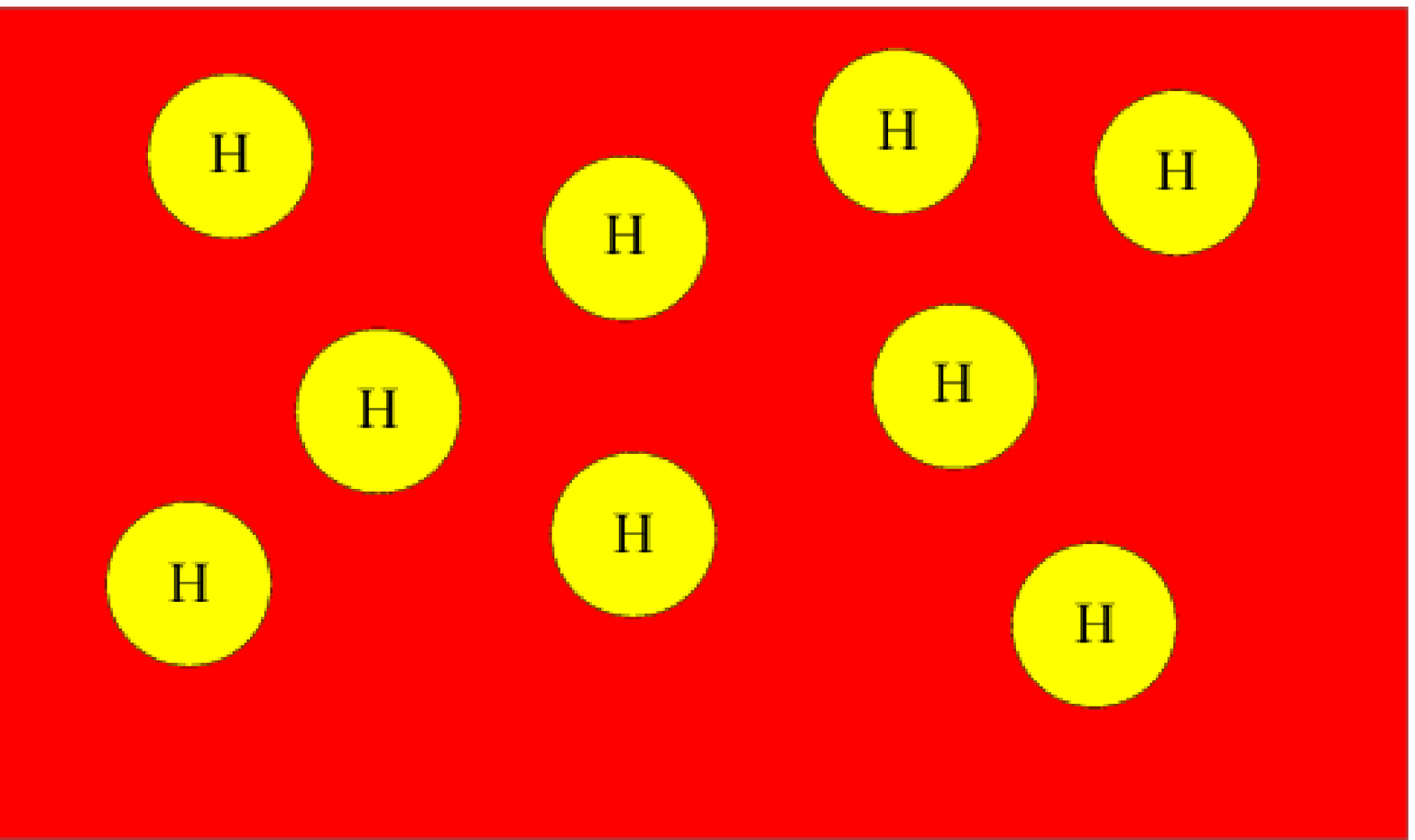}}
\parbox{1.4in}
{\hspace*{.4em}
\vspace*{-0in}
\includegraphics[angle=0,
width=1.\linewidth]{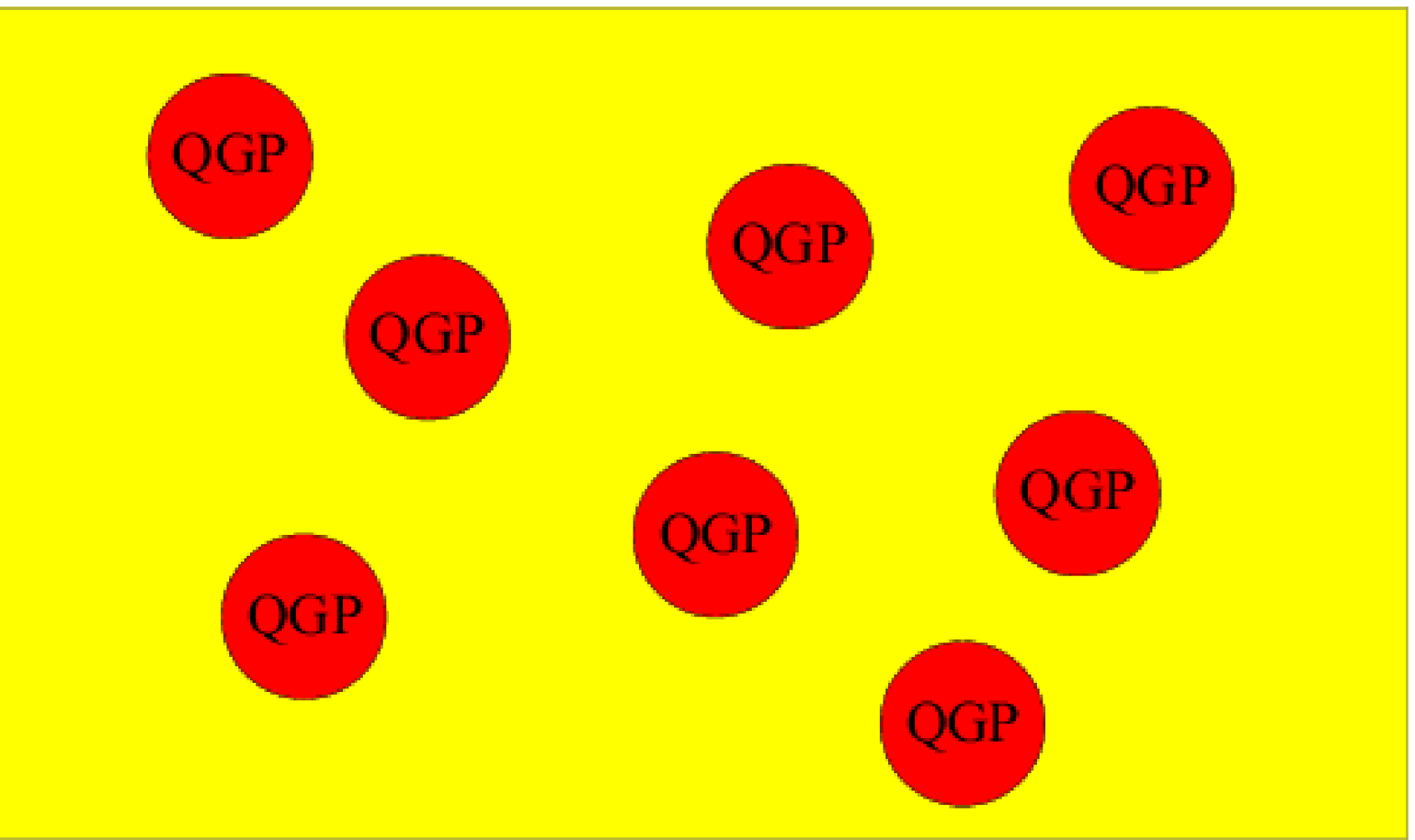}\\
\vspace*{-1.2in}
\hspace*{.4em}
\includegraphics[angle=0,
width=1.\linewidth]{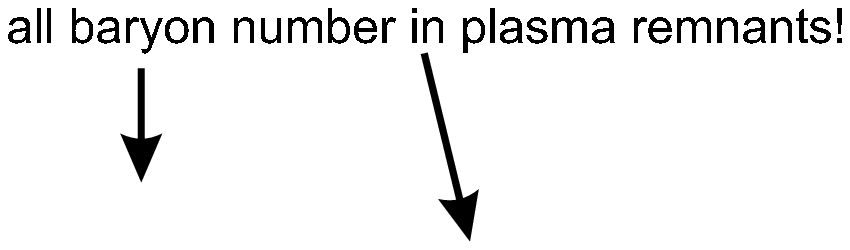}\\[.5ex]
}
\caption{Phase bubbles in the region of the confinement phase
transition, after~\cite{kam00}.}
\label{bubbles}
\end{figure}

\begin{figure}
{\includegraphics[angle=0,
width=.37\linewidth]{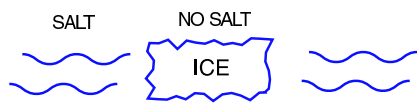}}
\caption{Seawater analogy.}
\label{sea}
\end{figure}

Associated with a first-order phase transition is the surface
tension~$\sigma$ and a possibility of supercooling.  For
sufficiently high $\sigma$, the early Universe might supercool
down to temperatures as low as half of the critical temperature
$T_c$, cf.\ Fig.~\ref{Ts}.  The large surface tension would
lead to a wide separation, by as much as $\ell \sim 1$m, of
the
forming hadronic bubbles and, eventually, as the hadronic
bubbles
grow and begin to fill all space, of the remnant quark-gluon
bubbles, cf.\ Fig.~\ref{bubbles}.  The separation would produce
large
nonuniformities, characterized by masses $M \sim 10^{18}$~kg
(i.e.\ of a medium size asteroid), in the distribution of the
baryon number following the hadronization, with the baryon
number concentrated in the regions that hadronize last.  The
excess baryon number would get trapped in the quark-gluon
bubbles, because the baryon number costs little in the
quark-gluon phase, with quarks being massless, and a lot in the
hadronic phase, with massive baryons.  An analogous situation
takes place when seawater freezes.  Then the salt appears in
the areas that are last to freeze, Fig.~\ref{sea}.
There are some cautioning theoretical and experimental
indications, though, regarding the scenario, that the surface
tension might not be very large between
in the quark-gluon and hadron phases.

\subsection{Supernova Explosions}

Type II supernova explosions are the source of at least half of
the nuclei heavier than iron around us.  Only very massive
stars, of masses $M \gtrsim 8 \, M_\odot$,
explode.  Generally,
the more massive a star, the shorter it lives, burning faster
due to higher density and temperature in its interior.
A star starts out burning hydrogen, then helium and
successively heavier nuclei; at each stage the products are
accumulated.
After a given fuel runs out, the gravitation compresses the
star core raising temperature and the next fuel ignites with
its burning preventing further compression.
When the core consists of iron only, the burning stops.  It
is then up to the
electron pressure (such as resisting the compression of solids)
to prevent the gravitational collapse of the core.
However, the electron pressure fails when the core exceeds the
threshold Chandrasekhar mass.  This is seen by examining the
contributions to the energy from gravity and from an
ultrarelativistic electron gas:
\bea
\nonumber
E = - \frac{3}{5} G  \frac{(N \, m_N)^2}{R} + N_e
\left( \frac{3}{4}  p_F  c + \frac{3  m_e^2  c^4}{ 2
p_F  c} + \ldots \right) + \ldots\\
= \frac{1}{R} \left( - \frac{3}{5}  G (N m_N)^2 +
\frac{3}{4} \hbar c
\left(\frac{9 \pi}{4}\right)^{1/3} N_e^{4/3}\right) + {\cal
O}(R) \, .
\eea
The electron Fermi momentum is proportional to the cube root of
electron density and, thus, is inversely proportional to core
radius, $p_F \propto \rho_e^{1/3} \propto 1/R$.  Both the
gravitational and electron energies are then inversely
proportional to the radius, but the electron energy grows only
as the number of electrons to the 4/3 power while the
gravitational energy as the square power of the nucleon number.
For the electron number equal to half the nucleon number, $N_e
= N_N/2$, the gravity wins over electrons for core mass
\beq
M > M_{th} = \left( \frac{5}{6} \frac{\hbar c}{G} \right)^{3/2}
\frac{3 \pi^{1/2} }{m_N^2} \sim 1.5 \, M_\odot  \, .
\eeq

\begin{figure}
{\includegraphics[angle=0,
width=.60\linewidth]{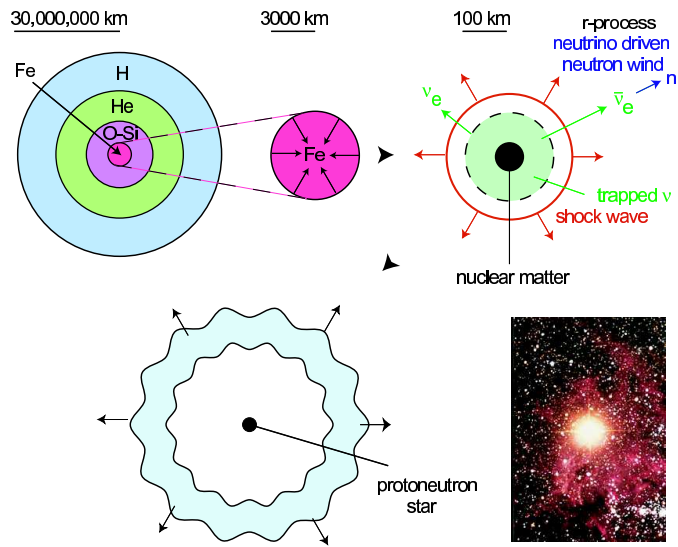}}
\caption{Supernova explosion.}
\label{SN}
\end{figure}

When the iron core exceeds the threshold mass, a
gravitational collapse of the core starts and progresses till
the nuclear densities are reached.  The nuclear matter is more
incompressible than the electron gas -- what starts as an
implosion gets reversed at the nuclear densities into an
explosion.  From the center of star a shock wave moves out, see
the schematic view in Fig.~\ref{SN}, while at the center a
so-called protoneutron star forms at a density of the order of
that in nuclei.  Inside, as the electron Fermi energy exceeds
the proton-neutron mass difference, the process of
neutronization takes place,
$e^- + p
\rightarrow  \nu_e + n$.  Additionally, thermal neutrinos are
copiously produced.  In the meantime, the shock moving through
the infalling material stalls outside of the protostar and
gets, most
likely, revived by the neutrinos coming out from the center.
Aside from propelling the shock,
the neutrinos drive the neutron wind from the center within
which copper, nickel, zinc and other elements form.
Eventually, the shock reaches the star surface producing a
magnificant display in the sky and throwing 7~$M_\odot$ of
material space.  The properties of nuclear matter,
where the
collapse reverses and that is the site of neutrino production,
are, however, generally not well known.

\subsection{Neutron Stars}

\begin{figure}
{\includegraphics[angle=0,height=1.85in,
width=.5\linewidth]{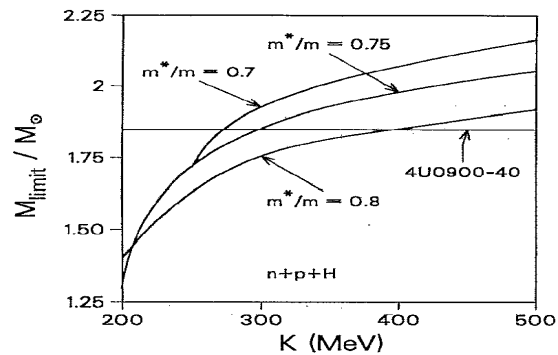}}
\caption{Limiting neutron star mass as a function
of the compression modulus of the corresponding
symmetric matter~\protect\cite{gle88}$^*$.}
\label{glen}
\end{figure}

\begin{figure}
{\includegraphics[angle=0,  
width=.63\linewidth]{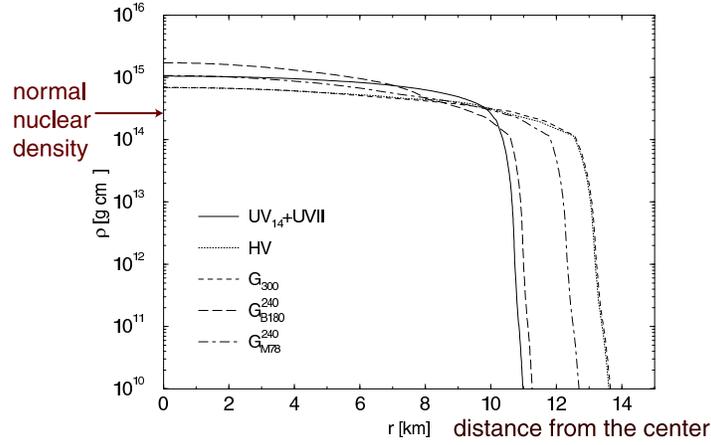}}
\caption{Density profile of a neutron star of mass $M=1.4 \,
M_\odot$, after~\protect\cite{sch96}.}
\label{ns}
\end{figure}

\begin{figure}
{\includegraphics[angle=0,height=2.5in,
width=.55\linewidth]{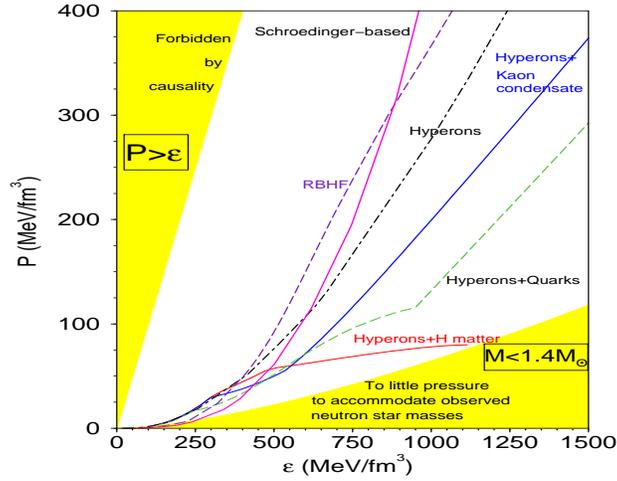}}
\caption{Pressure-energy relations~\protect\cite{web99} for
nuclear matter, employed in astrophysical calculations.}
\label{epeos}
\end{figure}

The protoneutron star eventually turns into a black hole or
into a neutron star.  Which is the case depends on the
properties of nuclear matter,
Fig.~\ref{glen}.  Dependent on those properties
are also the characteristics of the forming neutron star and,
in
particular, the density profile and radius, see Fig.~\ref{ns}.
In astrophysical modelling of neutron stars or of supernova
explosions, a host of nuclear EOS is employed,
such as those
in Fig.~\ref{epeos}, in terms of the dependence on pressure on
energy density.  Some EOS are excluded by causality (those with
high~$p$) and some by known masses of existing neutron stars
(those with low~$p$).  This still leaves a wide range of
possibilities; there are EOS taken from nonrelativistic and
relativistic calculations and some of the EOS incorporate
different types of phase transitions.

\begin{figure}
\parbox{2.20in}
{\vspace*{-.14in}
\includegraphics[angle=0,
width=1.\linewidth]{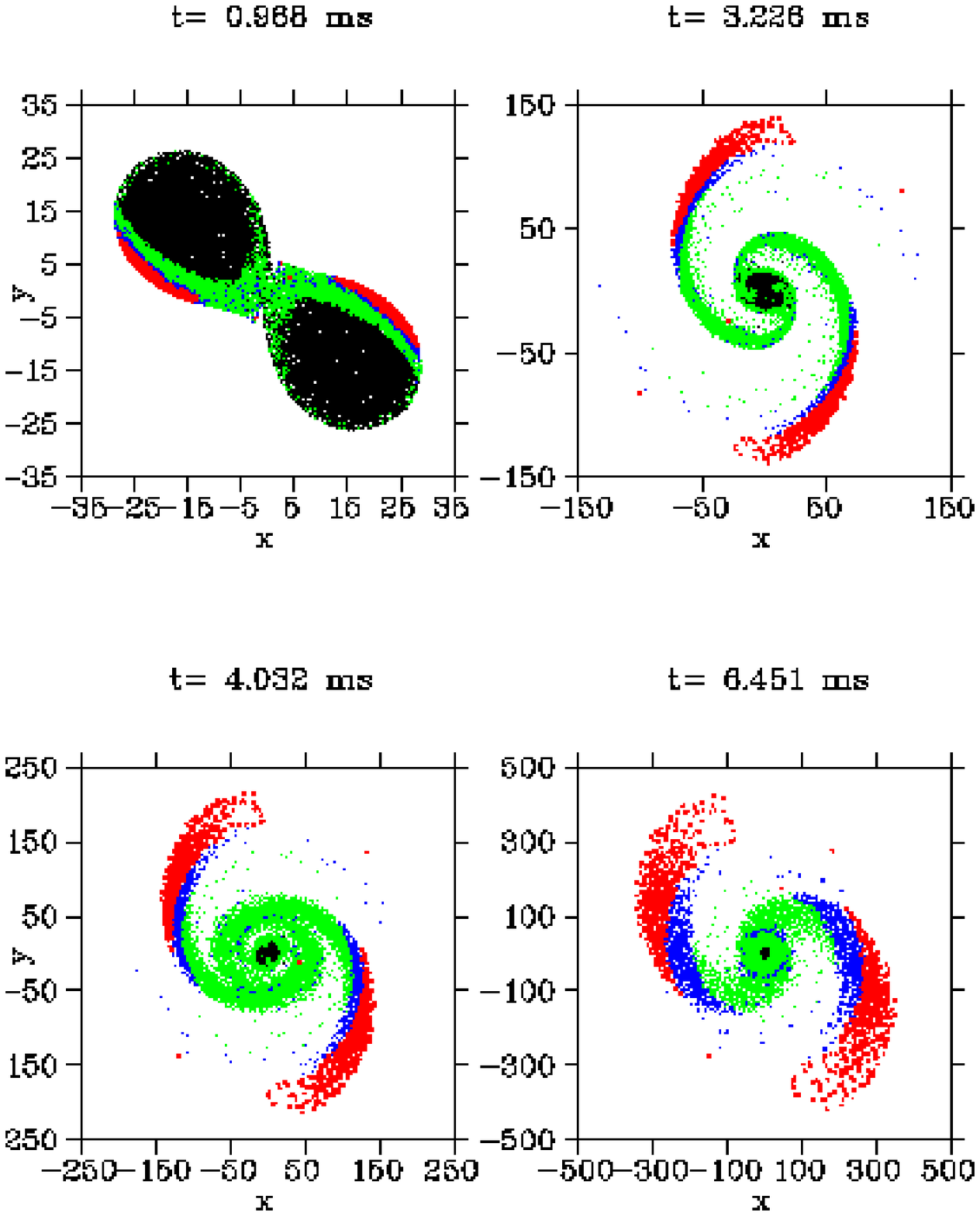}}
\parbox{2.2in}
{\hspace*{.8em}\includegraphics[angle=0,
width=1.\linewidth]{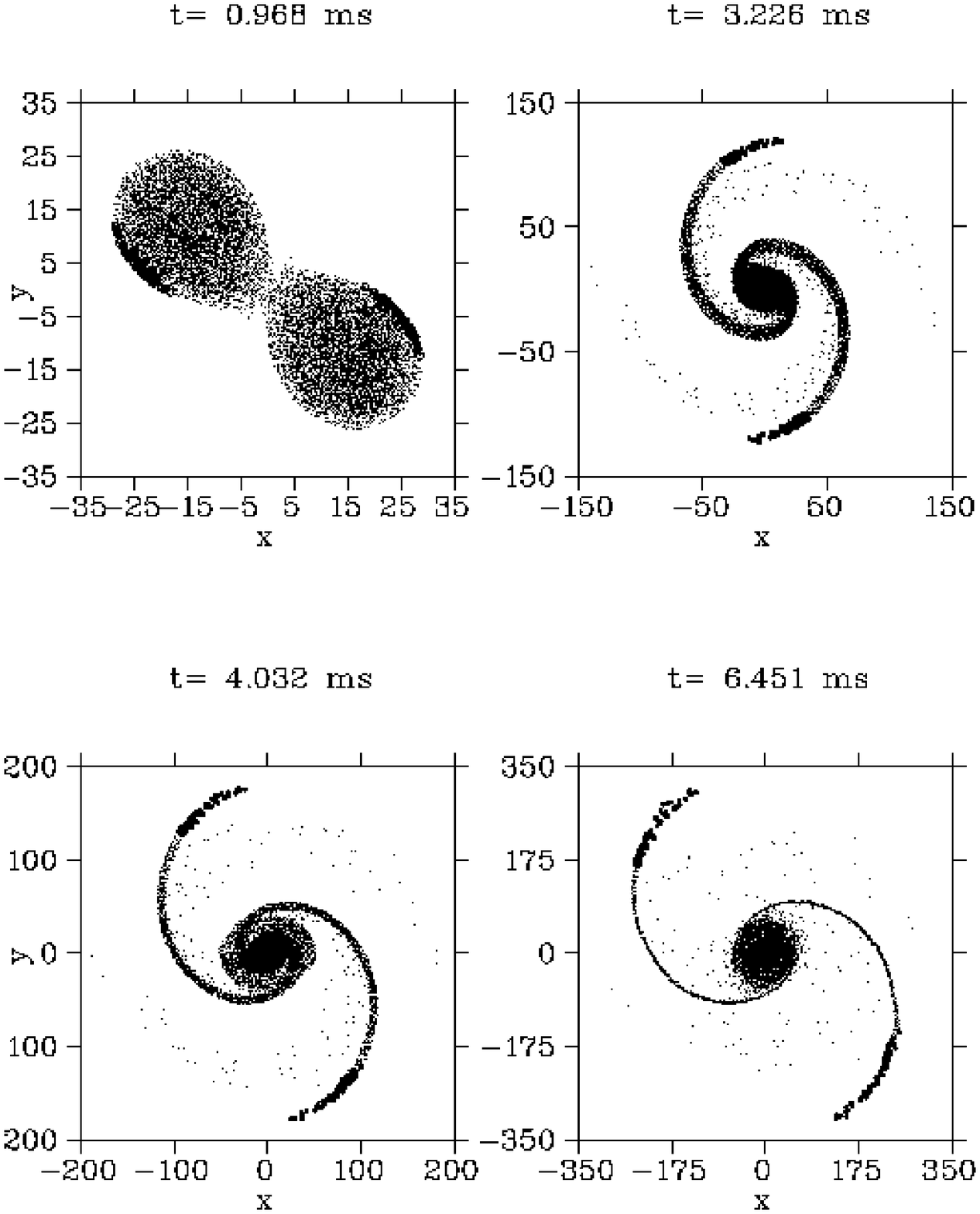}\\[-1ex]}
\caption{Neutron star mergers for soft (left panels) and
stiff (right panels) nuclear EOS, after~\protect\cite{ros99}.}
\label{ros99}
\end{figure}

A possible site for the synthesis of heavy elements,
other
than supernova explosions, are mergers of neutron stars.  These
mergers shed much more matter into space if the nuclear EOS is
relatively soft than when it is stiff,
Fig.~\ref{ros99}.

\section{Elementary Features of the Nuclear EOS}

\subsection{Energy Minimum}

\begin{figure}
{\includegraphics[angle=0,
width=.4\linewidth]{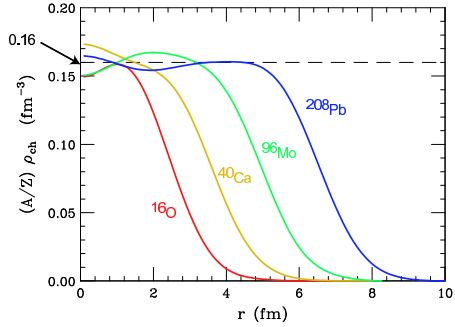}}
\caption{Nuclear densities deduced from electron scattering.}
\label{dens}
\end{figure}

\begin{figure}
{\includegraphics[angle=0, 
width=.55\linewidth]{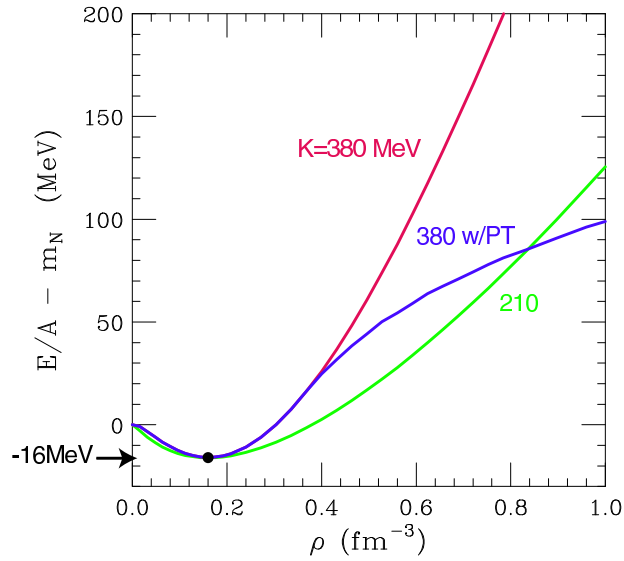}}
\caption{Energy per nucleon vs density in nuclear matter.}
\label{erho}
\end{figure}

The advances in the determination of the nuclear EOS have been,
generally, difficult.  The elementary information comes from
the Weizs\"acker binding-energy formula and from the
systematics
of nuclear density profiles.  The Weizs\"acker formula
separates out the contributions to the energy associated
with nuclear interactions and the interior and surface
of nuclei, the contributions associated with isospin
asymmetry and with Coulomb interactions, and the shell
correction, \bea
{-B(A,Z)} & = & {-16\, \mbox{MeV}} \, A + a_s \, A^{2/3}
\nonumber \\ &&
+ a_a \, \frac{(A
- 2\, Z)^2}{A}
+ a_c \, \frac{Z (Z-1)}{A^{1/3}}
- B_{p,s}  \, .
\eea
Nuclear densities, obtained from charge densities multiplied by
mass to charge number ratio, are seen to reach the same value,
$\rho_0 = 0.16$~fm$^{-3} \simeq
1/(6$~fm$^{3}$), for a wide range of nuclear masses, see
Fig.~\ref{dens}.  We conclude that the energy per nucleon in a
uniform symmetric nuclear matter at $T=0$, in the absence of
Coulomb interactions, has a minimum at the normal density
$\rho_0$ with
the energy value, relative to nucleon mass, of -16~MeV, from
the volume term in the binding formula, see Fig.~\ref{erho}.
As, obviously, the binding energy approaches zero for separated
nucleons at $\rho \rightarrow 0$, we actually know two
points in the ($T=0$)
dependence of the energy per nucleon, $E/A \equiv e/\rho$, on
density.

\begin{figure}
{\includegraphics[width=.11\linewidth]{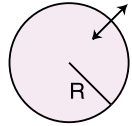}}
\caption{Oscillating nuclei were first considered as
sharp-edged spheres with the energy changing as a function of
the radius.}
\label{nucrad}
\end{figure}

The next nontrivial feature of the energy per nucleon is its
curvature in the dependence on $\rho$, around~$\rho_0$.  This
curvature is commonly quantified in terms of the so-called
nuclear incompressibility, with an unusal numerical factor:
\beq
K = 9 \, \rho_0^2 \, \frac{d^2}{d \rho^2}
\left(\frac{E}{A}\right)
= R^2 \, \frac{d^2}{dR^2} \left(\frac{E}{A}\right) \, .
\eeq
The factor stems from the fact that the nuclei were first
considered as sharp-edged spheres with the energy changing as a
function of the radius (Fig.~\ref{nucrad}).  To get an idea of
what might be expected for the incompressibility, one might
just run a parabola through the two known points on the curve
of $\frac{E}{A}(\rho)$.  The then resulting incompressibility
has a value of $K \sim 290$~MeV.  If the actual
incompressibility turns out to be below this benchmark
value, we may consider the nuclear EOS to be soft, and stiff if
the opposite is the case.

\subsection{Microscopic Calculations}

\begin{figure}
\parbox{2.6in}{
\hspace*{-1em}
\includegraphics[width=1.1\linewidth]{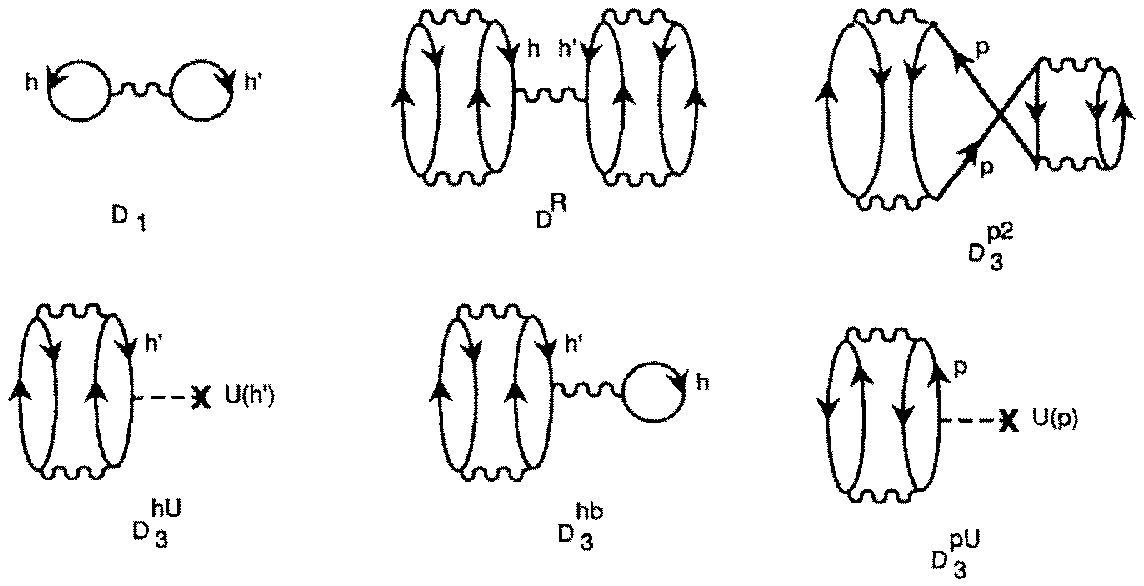}
}
\parbox{2.8in}{\hspace*{.5em}
\includegraphics[width=\linewidth,height=1.8in]{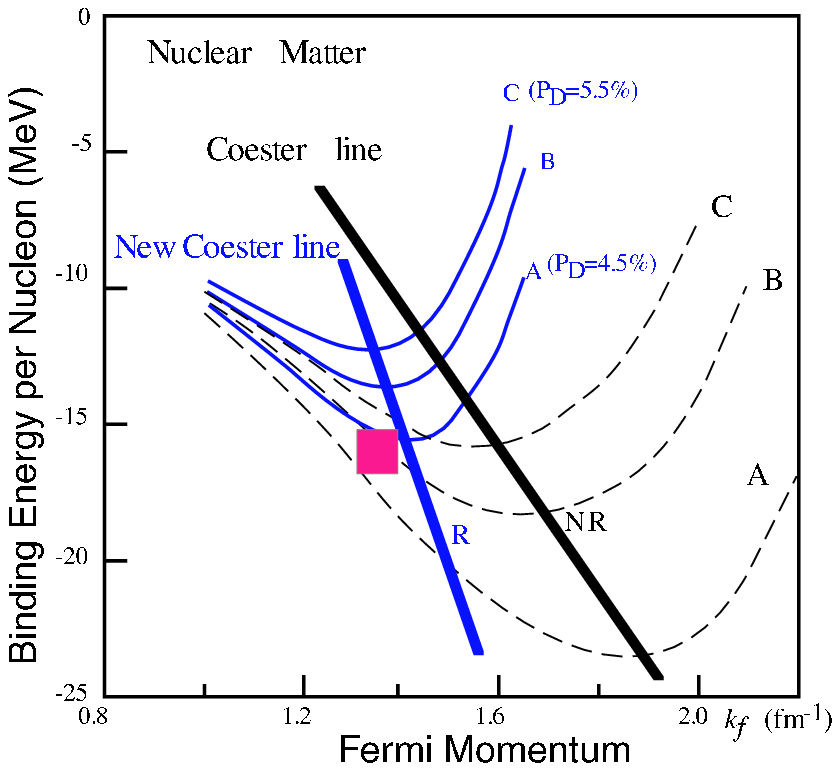} }
\caption{Left: Diagrams for different terms in the energy
per nucleon in many-body calculations~\protect\cite{bal90}.
Right:
Binding energy vs Fermi momentum in many-body calculations,
after~\cite{bro90}.}
\label{coester}
\end{figure}

To get the features of the nuclear EOS outside of the minimum,
one might turn to microscopic calculations, such as within
Brueckner and variational frameworks.  These
calculations
utilize elementary nucleon-nucleon interactions constrained by
nucleon-nucleon interactions and by deuteron properties.
However,
the nonrelativistic calculations with only nucleon-nucleon
interaction miss the known position of the minimum in the
nuclear EOS; the minimae line up along the so-called Coester
line (Fig.~\ref{coester}) in the energy vs density
or Fermi momentum, with the change of the version of
the interaction.  The relativistic calculations line up along
another Coester line that passes closer to the true minimum;
aside from relativity, though, those calculations are generally
more primitive than the nonrelativistic ones.

\begin{figure}
\includegraphics[width=.5\linewidth,height=2.05in]{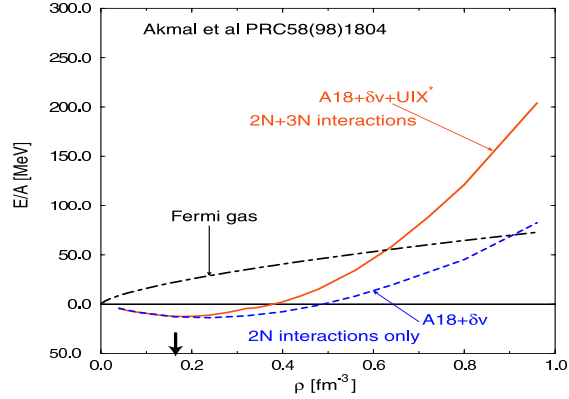}
\caption{Energy per nucleon in nuclear matter as a function of
density,
from a variational calculation of Ref.~\protect\cite{akm98}
with two- and three-nucleon interactions.}
\label{akmal}
\end{figure}

To get the right position of the minimum in the EOS,
Fig.~\ref{akmal}, it is necessary to incorporate three-nucleon
interactions in the microscopic calculations.  These
interactions are not well constrained by scattering, hampering
the predictive power of the theory.  In this situation, one may
want to turn to experiment to get the information on the EOS
away from the normal density.

\section{Incompressibility - Getting out of the Minimum}

\begin{figure}
\includegraphics[width=.4\linewidth]{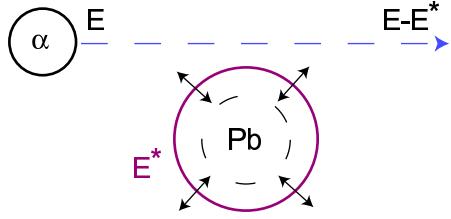}
\caption{Volume oscillations induced by alpha scattering.}
\label{alpha}
\end{figure}

The simplest way to determine the incompressibility
experimentally may seem to induce volume oscillations in a
nucleus.  This could be done by scattering $\alpha$ particles
off
a nucleus, Fig.~\ref{alpha}.  For the lowest excitation, the
excitation energy $E^*$, deduced from the final $\alpha$
energy, would be related to the classical frequency through
$E^*
= \hbar \Omega$, and the latter would be related to~$K$.
Let us examine the classical energy of an oscillating nucleus:
\bea
E_{tot} & = & \int d{\bf r} \, \rho \, \frac{m_N \, v^2}{2} +
\frac{1}{2} \, A \, K \, (R - R_0)^2\nonumber \\
& = & \frac{A m_N  \langle r^2 \rangle_A
\dot{R}^2}{2} + \frac{1}{2} \, A \, K \, (R - R_0)^2 \, ,
\eea
where we use the fact that, for a nucleus uniformly changing
its density, the velocity is proportional to the radius,
$v = \dot{R} \, (r/R)$.  We then obtain the energy of a simple
harmonic oscillator; the frequency is a square root of the
spring constant divided by mass constant, yielding:
\bea
E^* = \hbar \, \sqrt{\frac{K}{m_N \, \langle r^2 \rangle_A}}
\, .
\eea

There are complications regarding this
reasoning.  Thus, the nucleus is not a sharp-edged sphere
and the Coulomb interactions play a role in the oscillations
as well as nuclear interactions.  These effects may be
taken
care of by using an incompressibility constant characteristic
for a nucleus, $K \rightarrow K_A$, and isolating different
contributions in an analogy to those for the binding energy:
\beq
K_A = K + K_s \, A^{-1/3} + K_a \, \left(
\frac{N-Z}{A}
\right)^2 + K_c \, \frac{Z(Z-1)}{A^{4/3}} + \ldots  \, .
\eeq
With the corrections, it turns out that the
incompressibilities for
medium to heavy nuclei are about 2/3 of the incompressibility
for infinite nuclear matter,
e.g.\ $K^{Pb} \sim 0.64 \, K$; $K^{Sm} \simeq 0.67
\, K$.

However, there are more problems.  Thus, the density
oscillations lie high up in the excitation energy and get
broadened up.  This
may be remedied by employing a sum rule (notably, sum rules are
often robust tools in helping to link simple classical
considerations with the characteristics of quantum states):
\beq
\hbar \, \sqrt{\frac{K_A}{m_N \, \langle r^2 \rangle_A}}
= \sqrt{\frac{\langle E^{*3} \rangle_{0^+ \, spectrum}}
{\langle E^* \rangle_{0^+ \, spectrum}}} \, ,
\eeq
i.e.\ the incompressibility may be obtained from dividing the
third by the first moment of the spectrum.  An alternative is
to use a microscopic theory, with an effective interaction, to
describe both the excitation spectrum and the incompressibility
for infinite matter.

\begin{figure}
\includegraphics[width=.55\linewidth]{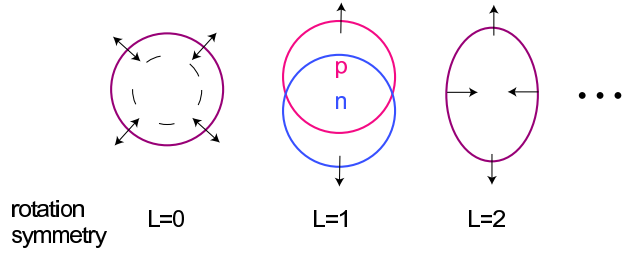}
\caption{Different collective oscillations transform
differently under rotations.}
\label{oscillations}
\end{figure}

\begin{figure}
\includegraphics[width=.37\linewidth]{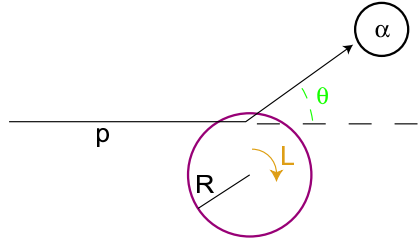}
\caption{Delivering angular momentum to a target.}
\label{falpha}
\end{figure}

The final complication is that other types of oscillations,
than that changing the density, are excited in scattering, such
as the oscillation of protons vs neutrons and the quadrupole
shape
oscillation, cf.~Fig.~\ref{oscillations}.  However, those
oscillations transform differently under rotations and,
correspondingly, the elementary excitations for those
oscillations are characterized by different angular momenta,
with the uniform density changes characterized by $L=0$.  It is
possible to isolate the $L=0$ excitations by analyzing
scattering at the very forward angles, Fig.~\ref{falpha}.
When the alpha particle scatters off a nucleus it transfers
linear and angular momenta to the nucleus.  The angular
momentum is limited by the product of the linear momentum
transfer and the distance over which the transfer occurs, i.e.\
roughly the sum of projectile and target radii.  At high beam
energies and small angles we get
\beq
L < |p - p'|\,R \approx p \, \theta \, R  \, .
\eeq
Excitations characterized by $L \ge 1 \, \hbar$ may suppressed
by looking at scattering into the angles $\theta <
\frac{\hbar}{pR}$, i.e.\ within the first diffraction peak.

\begin{figure}
\includegraphics[height=2.75in,width=.62\linewidth]{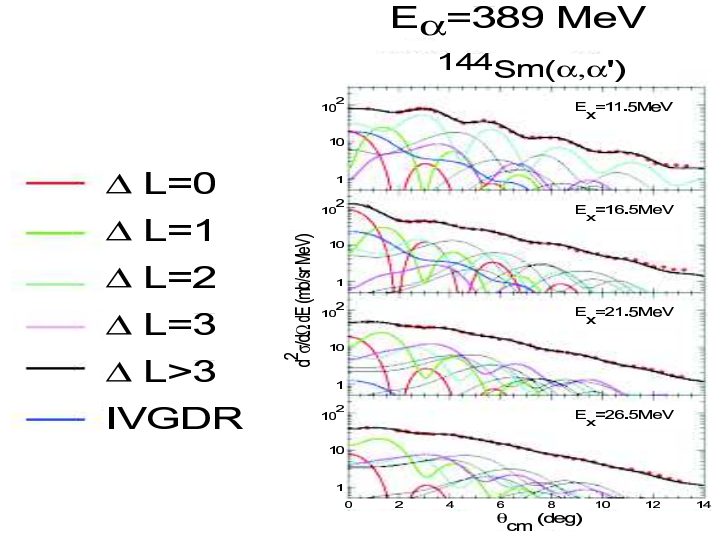}
\caption{Alpha scattering cross section from $^{144}$Sm,
determined in Ref.~\protect\cite{ito01}.}
\label{sm}
\end{figure}

\begin{figure}
\includegraphics[height=1.8in,width=2.2in]{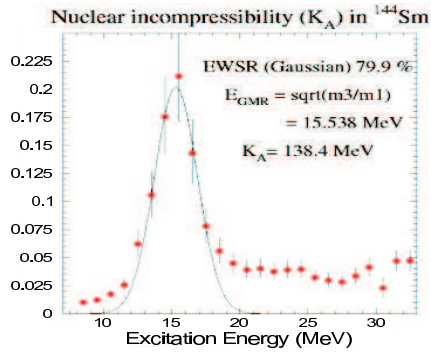}
\caption{$0^+$-strength function in $^{144}$Sm, determined in
Ref.~\protect\cite{ito01}.}
\label{sme}
\end{figure}

Scattering of alpha particles from different targets has been
carefully studied in recent years simultaneously as a function
of excitation energy and scattering angle,
allowing to isolate the contributions of $L=0$ excitations
\cite{you01,ito01}, see
Fig.~\ref{sm} for data from a samarium target.  For the shown
excitation energy of 16.5~MeV, a pronounced $L=0$ peak is
evident at low scattering angles.  The $L=0$ excitation
strength
is next shown for the samarium target in Fig.~\ref{sme}.
A peak is evident at the excitation energy of 15.5~MeV,
yielding an incompressibility of samarium
$
K^{Sm} = \frac{E^{*2} m_N \langle r^2
\rangle_A}{\hbar^2}
= 138 \, \mbox{MeV}$, and of nuclear matter
$
K = K^{Sm}/0.67  \sim 210 \, \mbox{MeV}
$.  However, explorations with microscopic models produce
different results for $K_A/K$.  In particular, relativistic
models can yield results in the range $K \sim (250 - 270)$~MeV
\cite{vre97}. Generally, the results are, though, on the soft
side of the incompressibility.

\section{EOS at Supranormal Densities from Flow}

Features of EOS at supranormal densities can be inferred from
flow produced in collisions of heavy nuclei at high
energies.  At low impact parameters, in those collisions,
macroscopic regions of high density are formed.  The collective
flow, that can be quantitatively assessed in collisions, is the
particle motion characterized by space-momentum correlations of
dynamic origin.  The flow can provide information on the
pressure generated in the collision.


To see how the flow relates to pressure, we may look at the
hydrodynamic Euler equation for the nuclear fluid, an analog of
the Newton equation, in a local frame where the collective
velocity vanishes, $v=0$:
\beq
(e + p) \, \frac{\partial}{\partial t} \, \vec{v} =
- \vec{\nabla} p \, .
\eeq
The collective velocity becomes an observable at the end of the
reaction.  In comparing to the Newton equation, we see that the
pressure
$p= \rho^2
\frac{\partial(e/\rho)}{\partial \rho}|_{s/\rho}$ plays the
role
of a potential for the hydrodynamic motion, while the density
of enthalpy $w=e+p$ plays the role of a mass.  In fact, at
moderate energies, the enthalpy density is practically the mass
density, $w \approx \rho \, m_N$.  We see from the Euler
equation that the collective flow can tell us about the
pressure
in comparison to enthalpy.  In establishing the relation, we
need
to know the spatial size where the pressure gradients develop
and this will determined by the nuclear size.  However, we also
need the time during the hydrodynamic motion develops and this
can represent a problem.


The equilibrium required for hydrodynamics is not quite
achieved in reactions and, thus, transport theory is actually
required
to establish links between the EOS and observables;
the hydrodynamics just yields important insights.  The reacting
system in the transport theory relying on Boltzmann equation is
described in terms of the phase-space distribution functions
$f$
for different particles.  In particular, the system energy is a
functional of the distributions, $E\lbrace f \rbrace$, and can
be parametrized to yield different EOS in
equilibrium.  The distributions follow a set of the Boltzmann
equations with single-particle energies that are functional
derivatives of the energy, $\epsilon = \delta E / \delta
f$:
\beq
{\partial f \over \partial t} + {\partial \epsilon \over
\partial {\bf p}} \, {\partial f \over \partial
{\bf r}} - {\partial \epsilon
\over \partial {\bf r}} \, {\partial f \over
\partial
{\bf p}} = I \, ,
\eeq
where $I$ is the collision integral.


The first observable that one may want to consider
to extract the information on EOS is the net radial or
transverse
collective energy. That energy may reach as much as half of the
total kinetic energy in a reaction.  Despite its magnitude, the
energy is not useful for extracting the information on EOS
because of the lack of information on how long
the energy develops.  Large pressures acting over a short
time
can produce the same net collective energy as low pressures
acting
over a long time.  This makes appearent the need for a timer in
reactions.

\begin{figure}
\centerline{\includegraphics[angle=0,
width=.77\linewidth]{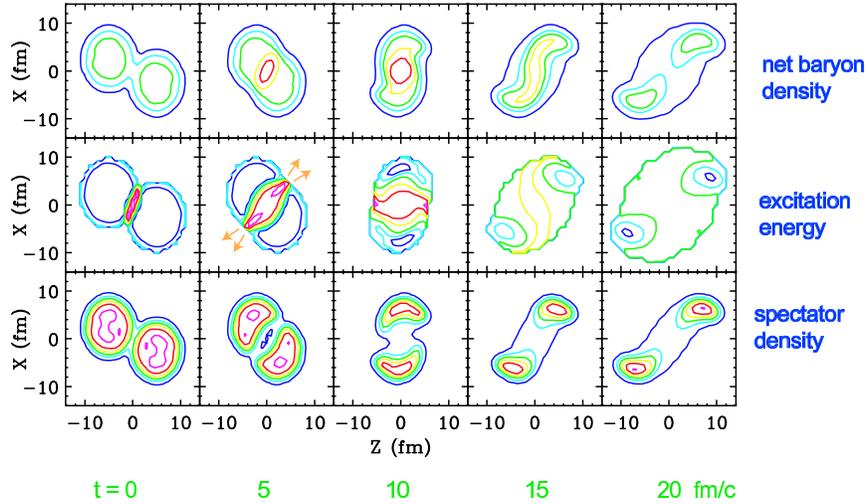}}
\caption{Reaction-plane contour plots for different quantities
in a $^{124}$Sn
+ $^{124}$Sn reaction at 800 MeV/nucleon and $b=6$~fm, from
transport simulations by Shi \protect\cite{shi01}.}
\label{contour}
\end{figure}

The role of the timer in reactions may be taken on by the
so-called spectators.  The spectator nucleons are those in the
periphery of an energetic reaction, weakly affected by the
reaction process, proceeding virtually at undisturbed original
velocity, see Fig.~\ref{contour}.  Participant
nucleons, on the other hand, are those closer to the center of
the reaction, participating in violent processes, subject to
matter compression and expansion in the reaction.  As the
participant zone expands, the spectators, moving at a
prescribed pace, shadow the expansion.  If the pressures in the
central region are high and the expansion is rapid, the
anisotropies
generated by the presence of spectators are going to be strong.
On the other hand, if the pressures are low and,
correspondingly, the expansion of the matter is
slow, the shadows left by spectators will not be very
pronounced.

There are different types of anisotropies in the emission that
the spectators can produce.  Thus, throughout the early stages
of
a collisions, the particles move primarily along the beam
axis in the center of mass.  However, during the compression
stage, the participants get locked within a channel, titled at
an angle, between the spectator pieces, cf.~Fig.~\ref{contour}.
As a consequence,
the forward and backward emitted particles acquire an
average deflection away from the beam axis, towards the
channel direction.  Another anisotropy may be
observed
for particles emitted in the transverse directions with zero
longitudinal velocity.  The region with compressed matter is
open to the vacuum in the direction perpendicular to the
reaction plane.  However, in the direction within the reaction
plane the region is shadowed by the participants.  Thus, more
particles are expected to be transversally emitted from the
participant
region perpendicular than within the direction plane.  The
anisotropy should be stronger the faster the expansion of the
compressed matter.

\begin{figure}
{\includegraphics[angle=0,height=2.3in,
width=.55\linewidth]{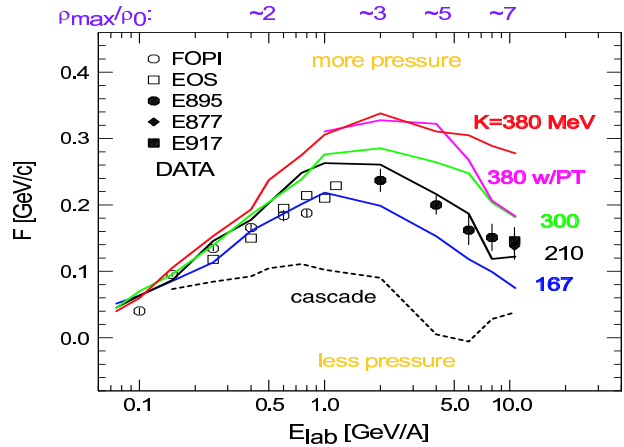}}
\caption{Sideward flow excitation function for Au + Au.  Data
and transport calculations are respresented, respectively, by
symbols and lines \protect\cite{dan01}.}
\label{flow}
\end{figure}

The different anisotropies have been quantified experimentally
over a wide range of bombarding energies.  Figure~\ref{flow}
shows the measure of the sideward forward-backward deflection
in Au + Au collisions as a function of the beam energy, with
symbols representing data.  Lines represent
simulations assuming different EOS.  On top of the figure,
typical maximal densities are indicated which are reached at a
given bombarding energy.
Without interaction
contributions to pressure, the simulations labelled
cascade produce far too weak anisotropies to be compatible with
data.  The simulations with EOS characterized by the
incompressibility $K=167$~MeV yield adequate anisotropy at
lower beam energies, but too low at higher
energies.  On the other hand, with the EOS characterized by
$K=380$~MeV, the anisotropy appears too high at virtually all
energies.  It should be mentioned that the incompressibilities
should be considered here as merely labels for the different
utilized EOS.  The
pressures resulting in the expansion are produced at densities
significantly higher than normal and, in fact, changing in the
course of the reaction.

\begin{figure}
{\includegraphics[angle=0,height=2.3in,
width=.55\linewidth]{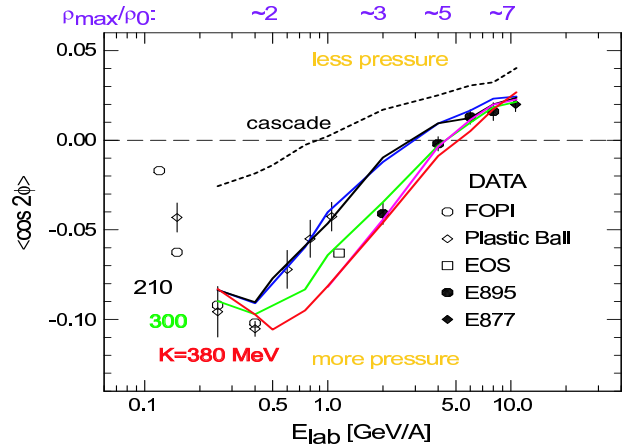}}
\caption{Elliptic flow excitation function for Au + Au.  Data
and transport calculations are respresented, respectively, by
symbols and lines \protect\cite{dan01}.}
\label{v2}
\end{figure}

\begin{figure}
\includegraphics[width=.3\linewidth,height=1.7in]{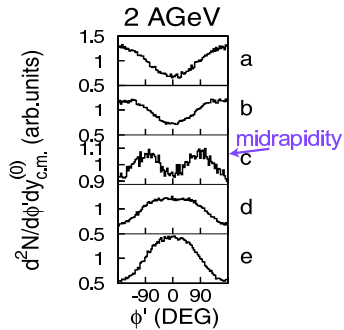}
\caption{Azimuthal distribution of protons from Au + Au
collisions at 2 GeV/nucleon in different rapidity intervals
\protect\cite{pin99}.}
\label{roymid}
\end{figure}

Figure~\ref{v2} shows next the anisotropy of emission at
midrapidity or zero longitudinal velocity in the c.m.,
cf.~Fig.~\ref{roymid}, with symbols representing data and
lines representing simulations.  Again, we see that without
interaction
contributions to pressure, simulations cannot reproduce the
measurements.  The simulations with $K=167$~MeV give too little
pressure at high energies, and those with $K=380$~MeV generally
too much.  A level of discrepancy is seen between data from
different experiments.

\begin{figure}
{\includegraphics[angle=0,
width=.54\linewidth]{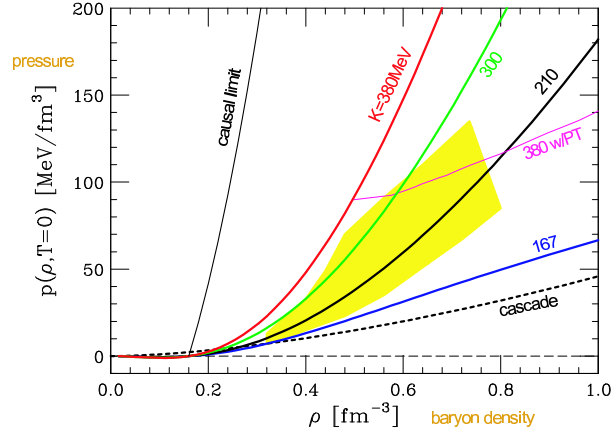}}
\caption{Constraints from flow on the $T=0$ pressure-density
relation, indicated
by the shaded region \protect\cite{dan01}.}
\label{Prho}
\end{figure}

\begin{figure}
{\includegraphics[angle=0,height=2.4in,
width=.54\linewidth]{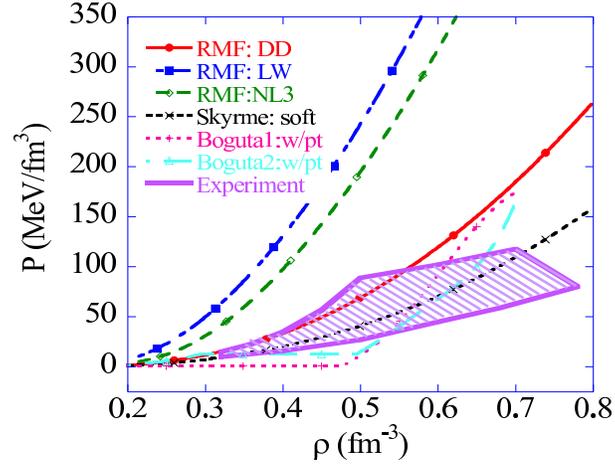}}
\caption{Impact of the constraints on models for EOS
\protect\cite{dan01}.}
\label{lynch}
\end{figure}

We see that no single EOS allows for a simultaneous description
of both types of anisotropies at all energies.  In particular,
the $K=210$~MeV EOS is best for the sideward anisotropy, and
the $K=300$~MeV EOS is the best for the other, so-called
elliptic, anisotropy.  We can use the discrepancy between the
conclusions drawn from the two types of anisotropies as a
measure of inaccuaracy of the theory and draw broad boundaries
on pressure as a function of density from what is
common in conclusions based on the two anisotropies.
To ensure
that the effects of compression dominate in the reaction over
other effects,
we limit
ourselves to densities higher than twice the normal.  The
boundaries on the pressure are shown in
Fig.~\ref{Prho} and they eliminate some of the more extreme
models for EOS utilized in nuclear physics, such as the
relativistic NL3 model and models assuming a phase transition
at relatively low densities, cf.~Fig.~\ref{lynch}.

\section{High-$T$ Low-$\rho$ Limits of the Hadronic World}

\begin{figure}
\includegraphics[width=.50\linewidth]{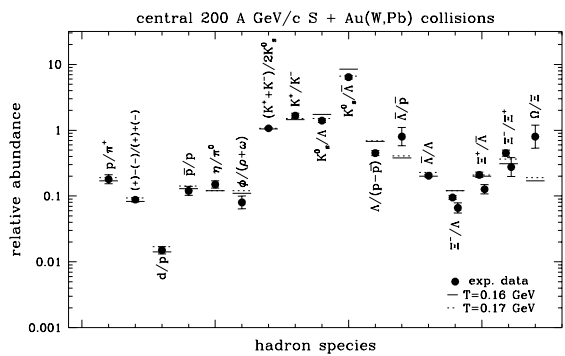}
\caption{Relative particle abundancies in measurements
(symbols) and calculated in the thermal freeze-out model
(lines) in Ref.~\protect\cite{bra96}.}
\label{abundance}
\end{figure}

\begin{figure}
\includegraphics[height=2.4in,width=.45\linewidth]{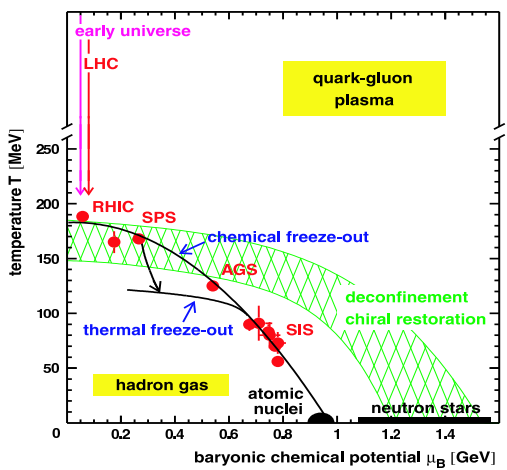}
\caption{Freeze-out temperature and baryon chemical potential$^*$.}
\label{stachel}
\end{figure}

In central reactions of medium to heavy nuclei, over a broad
range of bombarding energies, it is found that hadronic yields
are consistent with thermal equilibrium at definite $T$ and
$\mu$ when interactions appear to stop~\cite{bra96,cle99}.
This is illustrated
in Fig.~\ref{abundance} showing measured particle yields and
those calculated assuming thermal equilibrium.  The results
indicate that, at the deduced temperatures and chemical
potentials, the spectrum of hadrons is close to that in free
space and, thus, the phase transition to quark-gluon phase has
not been crossed.  The boundaries of the hadronic world, staked
out in this fashion, are shown in Fig.~\ref{stachel}.

\section{Conclusions and Outlook}

Nuclear EOS ties together different areas of physics.  Progress
on the EOS has been made in different directions.  Data on
giant monopole resonances (and also on giant vector resonances)
have
been collected with significant background reductions and high
resolution both in the energy and angle direction, allowing for
improved determinations of the nuclear incompressibility.
Anisotropies of flow from central reactions allow to constrain
the EOS at supranormal densities.  The parameters of freeze-out
in reactions allow to stake out the limits of the hadronic
world.  Additional sources of information on EOS that I had no
chance to talk about include measurements of neutron-star
properties, studies of nuclear systematics and lattice QCD
calculations.  Unconquered EOS frontiers include the dependence
of EOS on the isosopin degree of freedom and the detection of
the quark-gluon plasma.  The first frontier is, in particular,
to be tackled
at the NSCL coupled-cyclotrons and at the proposed RIA
accelerator.  In the baryonless regime, the second frontier is
pursued at RHIC.  However, the baryon-rich regime awaits
stepped-up dedicated studies with good resolution in bombarding
energy in the range of (2-40)~GeV/nucleon.

\begin{theacknowledgments}
Information provided by M.\ Itoh on giant resonances is
gratefully acknowledged.
This work was partially supported by the National Science
Foundation under Grant PHY-0070818.\\

$^*$Original figures from Refs.\ \cite{sta99} and \cite{sch96},
respectively, have been utilized, with permission from Elsevier
Science.
\end{theacknowledgments}

\end{document}